\documentclass[10pt,twocolumn]{article}
\usepackage[margin=0.75in]{geometry}
\usepackage{graphicx}
\usepackage{booktabs}
\usepackage{amsmath}
\usepackage{times}
\usepackage[hidelinks]{hyperref}
\usepackage{caption}
\usepackage{authblk}
\usepackage{xcolor}
\usepackage{balance}

\title{\textbf{Rendering on Real Silicon: GPU Render-Timing\\ as a Passive, AI-Resistant CAPTCHA Signal}}
\author[ ]{David Noever}
\author[ ]{Forrest McKee}
\affil[ ]{\textit{PeopleTec, Inc.}\\ \texttt{\{david.noever, forrest.mckee\}@peopletec.com}}
\date{}

\begin{document}
\maketitle

\begin{abstract}
Conventional CAPTCHAs pose puzzles that modern AI systems increasingly solve, while behavioral and cryptographic-attestation defenses carry privacy or enrollment costs. We investigate an orthogonal signal: the physical timing behavior of a client's GPU under a controlled WebGL rendering workload. Unlike WebGL \emph{fingerprinting}, which hashes pixel output into a static device identifier, we measure render-timing \emph{dynamics} to classify rather than identify, leaking no persistent identifier. We characterize the in-the-wild adversary with a 12-hour passive deployment (207 unsolicited requests; 86\% automated; 85\% of browser-claiming clients failed HTTP header-consistency checks). We then collect labeled GPU-timing samples through a single public endpoint exercised by real browsers (positive class, 13 distinct GPUs) and by keyed headless automation across a render-backend matrix (negative class). Software-rendered automation---empirically the dominant real-world adversary---separates from genuine GPUs by roughly $5\times$ in mean render time. On a confound-controlled comparison (identical GPU family and browser engine, differing only in headless vs.\ interactive execution), headless automation on real hardware still exhibits a distinct timing signature, separating from human samples by 75--106\% on frame jitter, timer-quantization ratio, and coefficient of variation. We report these as pilot-scale findings on a single GPU architecture and outline the cross-architecture collection required to establish generalization.
\end{abstract}

\begin{figure*}[t]
\centering
\includegraphics[width=0.86\textwidth]{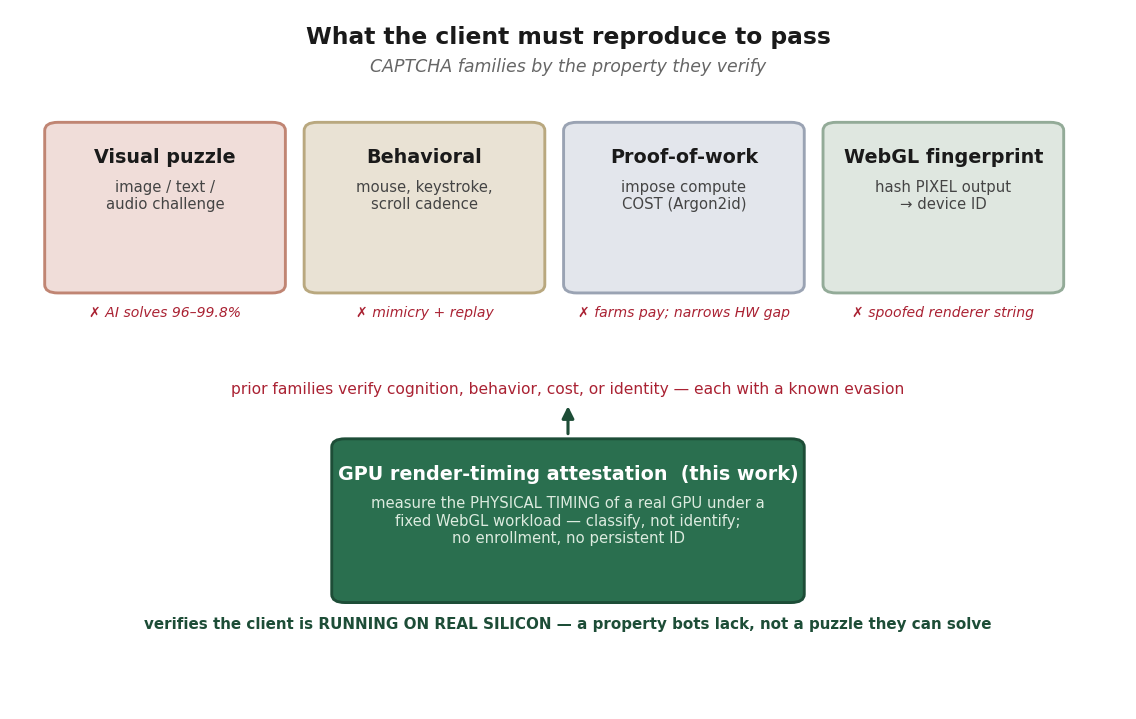}
\caption{CAPTCHA families organized by the property a client must reproduce to pass. Prior families verify cognition (visual puzzles), behavior (biometrics), computational cost (proof-of-work), or device identity (WebGL fingerprinting), each with a known evasion. The proposed approach verifies that the client is executing on genuine GPU hardware by measuring physical render timing---a property automated clients lack rather than a puzzle they can solve.}
\label{fig:overview}
\end{figure*}

\section{Introduction}
CAPTCHAs were designed around tasks easy for humans and hard for machines. That asymmetry has eroded: vision-language models (VLMs) solve image, text, and audio challenges with reported accuracies of 96--99.8\%~\cite{scopedesign2025,teoh2025}, CAPTCHA-solving services and human-in-the-loop APIs absorb the remainder, and even behavioral-biometric defenses face increasingly capable mimicry. The defensive frontier has shifted from posing harder puzzles toward verifying properties an automated client cannot cheaply reproduce.

This paper examines one such property: the physical timing behavior of a real GPU. We position the approach (Figure~\ref{fig:overview}) as a passive, physical-layer variant of hardware attestation. Where cryptographic attestation (TPM, WebAuthn, Secure Enclave) proves device identity through a signed challenge---at the cost of enrollment, a persistent identifier, and a trust chain that excludes many legitimate clients---a render-timing test asks only that the client \emph{behave like real silicon} while executing an ordinary graphics workload. No enrollment, no persistent identity, no key material.

We make three contributions:
\begin{enumerate}\itemsep2pt
\item An empirical characterization of the automated traffic reaching an exposed endpoint, establishing that the operative adversary is overwhelmingly render-free.
\item A single-endpoint methodology that gathers identically-measured timing samples from real browsers and from keyed headless automation, with trustworthy negative-class labels.
\item A pilot analysis showing clean separation of software-rendered bots, and a measurable, confound-controlled timing signature distinguishing even headless automation on real hardware, on one GPU architecture.
\end{enumerate}

\section{Related Work}
We organize prior work into five strands and state our relationship to each. Figure~\ref{fig:overview} summarizes the contrast.

\paragraph{Visual and cognitive CAPTCHAs, and AI solvers.}
The dominant CAPTCHA paradigm presents perceptual or cognitive challenges~\cite{guerar2021,trong2023}. Their security has collapsed against modern AI: generative and adversarial solvers defeat text and image challenges~\cite{ye2018}, and recent agentic VLM systems such as Halligan~\cite{teoh2025} generalize across visual challenge types, solving unseen CAPTCHAs by parsing an objective, abstracting the challenge, and synthesizing an interaction program. Notably, that threat model explicitly assumes the CAPTCHA is \emph{not} vulnerable to side channels and that the attacker operates ``vision-level only.'' Our method is \emph{complementary}: it adds a non-visual, physical signal that lies precisely in the gap such solvers exclude.

\paragraph{Behavioral biometrics.}
Production systems (reCAPTCHA v3, Cloudflare Turnstile, hCaptcha) score sessions on mouse micro-movements, scroll velocity, keystroke cadence, and timing regularity~\cite{scopedesign2025,knowledgesdk2026}, and the research literature has extended this with synthetic-trajectory and keystroke-generation defenses~\cite{acien2022,dealcala2023}. These observe \emph{human motor noise}. We observe \emph{hardware execution physics}; the two are orthogonal and combine naturally within a multi-signal risk score. A key practical difference: behavioral signals require interaction time and are vulnerable to replay and learned mimicry, whereas a render-timing probe completes in a fixed sub-second window and measures a property the adversary cannot supply without real hardware.

\paragraph{Proof-of-work CAPTCHAs.}
Proof-of-work (PoW) systems such as ALTCHA and Arkose Labs impose a client-side computational cost, shifting attack economics through cost asymmetry~\cite{altcha2024,arkose2026}. Critically, modern PoW deliberately uses memory-hard functions (Argon2id, scrypt) \emph{to narrow} the performance gap between commodity devices and specialized hardware~\cite{altcha2024}. Our approach inverts this: we use a rendering workload not as a cost toll but as a \emph{measurement probe}, and we \emph{exploit} rather than erase the hardware gap---the physical timing distribution of the GPU executing the probe is the signal.

\paragraph{WebGL fingerprinting.}
The closest-named prior art renders a hidden scene and hashes pixel output, combined with the \texttt{WEBGL\_debug\_renderer\_info} string, into a stable device identifier~\cite{spidra2026}. This is widely deployed for tracking and bot detection, and is recognized as harder to spoof than canvas or user-agent fingerprints because it derives from hardware behavior~\cite{spidra2026}. Two distinctions define our contribution. First, \emph{purpose}: fingerprinting performs \emph{identification} (a persistent per-device hash, with attendant privacy concerns and tracking use), whereas we perform \emph{classification} (human vs.\ automated) and retain no persistent identifier. Second, and decisively, \emph{signal}: the entire fingerprint-evasion ecosystem---antidetect browsers, canvas-noise injection, renderer-string spoofing, low-level C++ patching~\cite{browserless2025}---targets the \emph{pixel hash and renderer string}. None of it addresses render \emph{timing}, because timing is not what these systems fingerprint. Our signal is therefore orthogonal to, and survives, the deployed spoofing stack: a tool that perfectly forges the renderer string and pixel output of an ``NVIDIA RTX 4070'' still cannot forge the timing distribution of one it does not physically possess.

\paragraph{GPU timing side-channels.}
A peer-reviewed body of work establishes that GPU and browser timing carry exploitable hardware information. JavaScript microarchitectural attacks recover high-resolution timing despite browser timer coarsening~\cite{webgpuspy2024}, and WebGPU/WebGL timing has been used for cache attacks and website fingerprinting. These works use GPU timing \emph{offensively} (to extract secrets or identify activity). We repurpose the same physical phenomenon \emph{defensively}: the hardware-dependence of render timing that makes it a side-channel risk is precisely what makes it a bot-classification signal. This literature grounds our core assumption---that render timing reflects physical hardware---in established results.

\section{Threat Model: What Reaches the Endpoint}
Before measuring GPU timing, we characterized the adversary empirically. A public endpoint on commodity cloud infrastructure passively logged unsolicited traffic over a 12-hour window, recording for each request the source address, autonomous system (resolved via Team Cymru), requested path, User-Agent, and the HTTP headers a conforming browser emits.

The endpoint received 207 unsolicited requests; the first arrived 12.6\,s after launch. Although 192/207 (93\%) presented a browser-like User-Agent, only 28 (15\% of those) were consistent with a genuine browser once presence of \texttt{Sec-Fetch-*}, \texttt{Accept-Language}, and a content-specific \texttt{Accept} header was required---matching the header-based detection used in practice~\cite{knowledgesdk2026}. The remaining 85\% declared a browser identity while omitting headers no real browser omits. In aggregate, 86\% of requests were automated; requested paths were dominated by secret- and configuration-discovery probes; exactly one request touched a data-collection path and \emph{none} reached any rendering endpoint.

The operative finding: \textbf{the in-the-wild adversary is render-free}. These clients do not execute client-side JavaScript and never reach a rendering stage, so a render-gated challenge filters them by construction---but they also yield no timing data, so characterizing evasive bots that \emph{do} render requires the deliberate collection described next.

\section{Methodology}
\subsection{Single-endpoint collection}
A single public page serves a shared WebGL harness and accepts session records at one endpoint. Both classes flow through the same instrument, eliminating measurement confounds between them. Real visitors open the page and, with consent, post a timing record (positive class). Headless automation is driven to the same URL by a sweep tool carrying a server-side secret, so its label is trustworthy at collection (negative class). The server stamps each record with receive-side metadata the client cannot forge, including the header-consistency signal above.

\subsection{The harness}
The harness issues a deterministic, fragment-shader-bound WebGL workload and records per-frame render time across a fixed window after warm-up. Each frame forces pipeline completion via a synchronous pixel read-back, so measured time reflects GPU execution rather than asynchronous queue submission. The raw per-frame series is retained; features are derived offline so they can be revised without re-collection. The harness also records the unmasked renderer string, browser/platform metadata, and the observed timing-source resolution.

\subsection{Features}
Per session we derive distributional statistics: mean and median render time; dispersion (standard deviation, coefficient of variation [CV], inter-quartile range); shape (skewness, kurtosis); frame-to-frame jitter (mean absolute first difference); and a timer-quantization indicator (distinct timing values / frame count).

\subsection{Negative-class backend matrix}
The sweep navigates to the page across automation stacks and backends: Chromium over SwiftShader and SwANGLE (software), Chromium with GPU disabled (CPU fallback), headless Firefox and WebKit, and Chromium acquiring a physical GPU in headless mode. The last cell is the \emph{hard negative}---automation on real hardware---and prevents a classifier from succeeding merely by keying on a software-renderer string. Because headless Chromium frequently falls back to software, the page self-gates this cell, posting only sessions in which a genuine hardware renderer was acquired.

\subsection{Evaluation discipline}
Any classifier is evaluated under a \textbf{device-held-out} split: entire GPUs are assigned to train or test, never individual samples. Training on some GPUs and testing on GPUs unseen in training is the only honest measure of generalization; a per-sample split would leak device identity and inflate accuracy.

\subsection{Ethical considerations}
The threat-model measurement passively recorded metadata (source address, autonomous system, requested path, headers) of unsolicited requests arriving at infrastructure we control; no challenge or payload was returned beyond a static page, and no third-party service was probed. Positive-class timing samples were contributed by visitors who affirmatively consented on the collection page after disclosure of what is transmitted; records contain no cookies, account data, or persistent identifiers, consistent with the classification-not-identification design. Negative-class sessions were generated exclusively by automation we operated against our own endpoint.

\section{Results}
We group usable sessions into three classes: human (real GPU, interactive), software bot (software-rendered headless), and hard negative (real GPU, headless). Table~\ref{tab:summary} summarizes the pilot dataset.

\begin{table}[t]
\centering\small
\caption{Pilot dataset summary by class.}
\label{tab:summary}
\begin{tabular}{@{}lrrrr@{}}
\toprule
Group & $n$ & GPUs & mean (ms) & jitter \\
\midrule
Human (real, interactive) & 26 & 13 & 5.09 & 0.42 \\
Software bot & 273 & 4 & 26.28 & 2.03 \\
Hard negative (real, headless) & 20 & 2 & 8.25 & 0.91 \\
\bottomrule
\end{tabular}
\end{table}

\subsection{Software automation separates widely}
\begin{figure}[t]
\centering
\includegraphics[width=\columnwidth]{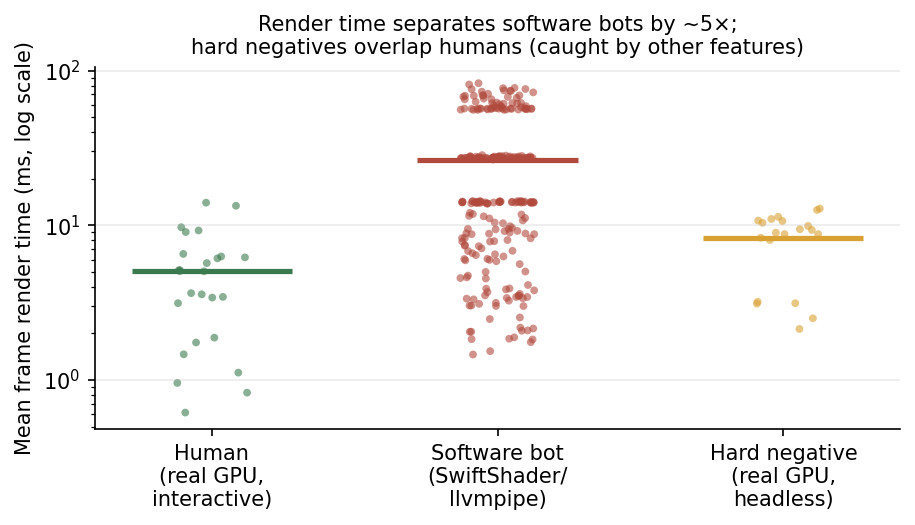}
\caption{Mean frame render time by class (log scale). Software bots are $\sim$5$\times$ slower than genuine GPUs; hard negatives overlap the human range on this feature and require other features to separate.}
\label{fig:rendertime}
\end{figure}

Software-rendered bots render the workload far more slowly than genuine GPUs---26.3\,ms vs.\ 5.1\,ms, a $5.2\times$ gap---with higher jitter and quantization (Figure~\ref{fig:rendertime}). As this population is also the entire in-the-wild adversary, a simple render-time threshold filters realistic automated traffic.

\subsection{Headless on real hardware leaves a signature}
\begin{figure}[t]
\centering
\includegraphics[width=\columnwidth]{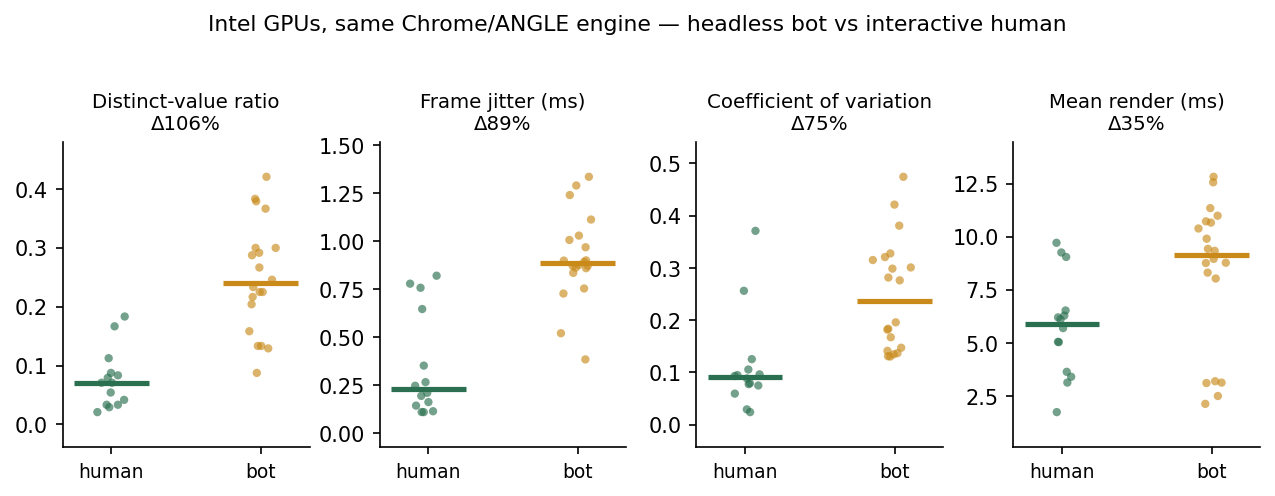}
\caption{Per-feature distributions, Intel GPUs under matched engine. Headless execution shows higher timer-quantization ratio, jitter, and CV than interactive use of the same hardware.}
\label{fig:intel}
\end{figure}

The harder question is whether automation on a genuine GPU is distinguishable from a human on the same hardware. Comparing across browser engines would confound the result, since engines differ in WebGL implementation and timer behavior. We therefore restrict to one confound-controlled condition: the same GPU family (Intel integrated) and engine (Chromium/ANGLE), differing only in headless vs.\ interactive execution (Table~\ref{tab:intel}, Figure~\ref{fig:intel}).

\begin{table}[t]
\centering\small
\caption{Confound-controlled comparison: Intel GPUs, Chromium/ANGLE, headless bot vs.\ interactive human.}
\label{tab:intel}
\begin{tabular}{@{}lrrr@{}}
\toprule
Feature & Human & Bot & Sep. \\
& ($n{=}14$) & ($n{=}20$) & \\
\midrule
distinct-value ratio & 0.076 & 0.249 & 106\% \\
frame jitter (ms) & 0.351 & 0.912 & 89\% \\
coeff.\ of variation & 0.112 & 0.247 & 75\% \\
mean render (ms) & 5.78 & 8.25 & 35\% \\
\bottomrule
\end{tabular}
\end{table}

\begin{figure}[t]
\centering
\includegraphics[width=\columnwidth]{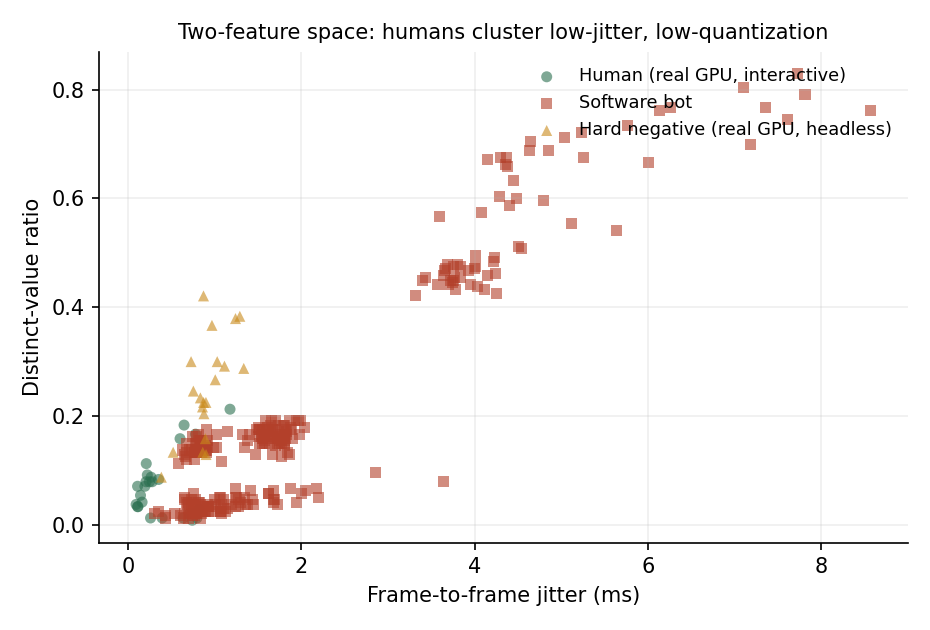}
\caption{Two-feature space (jitter vs.\ quantization ratio). Humans cluster at low jitter and low quantization; software bots spread across high jitter; hard negatives form a distinct band of elevated quantization at low jitter.}
\label{fig:space}
\end{figure}

Headless sessions show $3.3\times$ higher quantization ratio, $2.6\times$ higher jitter, and $2.2\times$ higher CV than interactive sessions on the same hardware. The direction is consistent with a physical mechanism: a headless browser lacks the display compositor and refresh pacing that regularize frame timing in interactive use, so it renders with \emph{more} timing roughness, not less---the opposite of the naive intuition that automation is flatter, and the reason a single CV threshold is insufficient (Figure~\ref{fig:space}).

\section{Limitations and Scope}
\textbf{Single architecture.} The confound-controlled signature is shown on Intel integrated GPUs only---the single family for which we hold both human and hard-negative samples. Whether it replicates on discrete NVIDIA/AMD and Apple Silicon is the central open question; the mechanism is architecture-independent in principle, but faster discrete GPUs may compress the timing differences below measurability. We make no cross-architecture claim.

\textbf{Pilot scale.} The positive class spans 13 GPUs / 26 sessions; the hard negative spans 2 GPUs / 20 sessions. These support separability in principle on the measured devices, not a trained discriminator with characterized error rates.

\textbf{Adversary adaptation.} An adversary can render on real hardware (raising cost, not enabling free evasion) or inject synthetic jitter to mimic interactive timing. The method is best understood as a low-friction, hard-to-spoof \emph{signal} within multi-signal risk assessment, and as a near-complete filter against the render-free automation that dominates real traffic.

\textbf{Timer hardening.} Browsers that coarsen or jitter high-resolution timers compress the signal; the harness records observed timer resolution per session, but the population-level impact is not yet quantified.

\section{Conclusion}
The in-the-wild automated adversary does not render, so a render-gated challenge filters it almost entirely. Beyond that, even headless automation on genuine hardware leaves a measurable timing signature relative to interactive use of the same hardware, at least on the Intel GPUs measured here. Establishing cross-architecture generalization is a matter of continued, architecture-stratified collection, for which the single-endpoint methodology described here is designed; the immediate targets are matched human/hard-negative pairs on discrete NVIDIA and AMD GPUs and on Apple Silicon.

\balance

\begin{thebibliography}{99}
\small
\bibitem{scopedesign2025} ScopeDesign. \emph{Why Can't Bots Solve CAPTCHA? Bot Detection Guide 2024--2025}, 2025.
\bibitem{teoh2025} J. Teoh et al. \emph{Halligan: A Generalized Visual CAPTCHA Solver via VLM Agents}. USENIX Security, 2025.
\bibitem{guerar2021} M. Guerar et al. Gotta CAPTCHA'em All: A Survey of 20 Years of the Human-or-Computer Dilemma. \emph{ACM Comput. Surv.} 54(9), 2021.
\bibitem{trong2023} N.D. Trong et al. New Cognitive Deep-Learning CAPTCHA. \emph{Sensors} 23(4):2338, 2023.
\bibitem{ye2018} G. Ye et al. Yet Another Text CAPTCHA Solver: A GAN-Based Approach. \emph{ACM CCS}, 2018.
\bibitem{knowledgesdk2026} KnowledgeSDK. \emph{Anti-Bot Detection in 2026}, 2026.
\bibitem{acien2022} A. Acien et al. BeCAPTCHA-mouse: Synthetic Mouse Trajectories and Improved Bot Detection. \emph{Pattern Recognition} 127:108643, 2022.
\bibitem{dealcala2023} D. DeAlcala et al. BeCAPTCHA-type: Biometric Keystroke Data Generation for Improved Bot Detection. \emph{CVPR Workshops}, 2023.
\bibitem{altcha2024} ALTCHA. \emph{Proof-of-Work CAPTCHA Documentation}, 2024. \url{https://altcha.org}.
\bibitem{arkose2026} Arkose Labs. \emph{Proof of Work: Invisible Economic Barrier}, 2026.
\bibitem{spidra2026} Spidra. \emph{WebGL Fingerprinting and How to Bypass It When Scraping}, 2026.
\bibitem{browserless2025} Browserless. \emph{Browser Fingerprinting Guide: Detection \& Bypass Methods}, 2025.
\bibitem{webgpuspy2024} Z. Wang et al. WebGPU-SPY: Finding Fingerprints in the Sandbox through GPU Cache Attacks. \emph{arXiv:2401.04349}, 2024.
\end{thebibliography}
\end{document}